\documentclass[showpacs, twocolumn]{revtex4-1}
\usepackage{graphicx}
\usepackage{amsmath}
\usepackage{appendix}
\begin{document}

\title{Dipolar Bose gas with three-body interactions in weak disorder}

\author{Redaouia Keltoum$^{1}$ and Abdel\^{a}ali Boudjem\^{a}a$^{1,2}$}

\affiliation{$^1$ Laboratory of Mechanics and Energy, \\
$^2$ Department of Physics, Faculty of Exact Sciences and Informatics, 
Hassiba Benbouali University of Chlef, P.O. Box 78, 02000, Chlef, Algeria.}

%\date{\today}

\begin{abstract}

We study effects of weak disorder with Gaussian correlation function on a dipolar Bose gas with three-body interactions using the Hartree-Fock-Bogoliubov theory.
Corrections due to quantum, thermal and disorder fluctuations to the condensate depletion, the one-body density correlation function 
as well as to the equation of state and the ground state energy are properly calculated.
We show that the intriguing interplay of the disorder, dipole-dipole interactions and three-body interactions plays a fundamental role in the physics of the system.
Interestingly, we find that the three-body interactions release atoms localized in the respective minima of the random potential.

\end{abstract}


\maketitle

\section{Introduction}

Since the first observation of a Bose-Einstein condensate (BEC) with dipole-dipole interactions (DDI) in 2005, 
quantum dilute atomic gases have attracted major attention both theoretically and experimentally \cite{Baranov, Pfau, Carr,Pupillo}. 
Dipolar BECs provide rich new physics, not encountered in systems with the contact interaction, thanks to the long-range and anisotropic character of the DDI.

The properties of dipolar BECs in weak random potential have recently sparked a great interest \cite{Krum,Nik, Ghab, Boudj,Boudj1,Boudj2, Boudj3}. 
It is widely agreed that the stability and the shape of such systems depend critically on the interplay between the DDI and the disorder potential. 
For instance, the superfluidity acquires a characteristic direction dependence due to the anisotropy of the DDI  \cite{Krum,Nik, Ghab, Boudj,Boudj1}.  
The studies of two-dimensional (2D) disordered dipolar gases may offer the possibility for the observation of the superglass state \cite{Boudj2, Boudj3}.
Most recently, impacts of the Lee-Huang-Yang (LHY) quantum corrections on a dirty dipolar Bose gas have been analyzed by one of us \cite{Boudj5} using a perturbative theory.
It is found that the LHY quantum fluctuations lead to reduce the disorder effects inside the condensate preventing the formation of the Bose glass state.

On the other hand, three-body interactions (TBI) play a key role in a wide variety of interesting physical phenomena, 
and provide a new physics not existed in systems with two-body interactions. 
Inelastic three-body processes, including observations of Efimov quantum states and atom loss from recombination have been reported in Refs \cite {Eff, Eff1, Bed, Kra, Brut}.
Weakly interacting Bose and Fermi gases with competing attractive two-body and large repulsive TBI may form droplets \cite {Bulg}. 
Effects of TBI in ultracold bosonic atoms loaded in an optical lattice or a superlattice were also studied in \cite{Daly1, Mazz, Singh, Mahm}. 
It was shown also that the TBI in Bose condensate may singnificantly modify the collective excitations \cite{Abdul, Hamid, Chen}, the transition temperature,
the condensate depletion and the stability of a BEC \cite {Peng, Mash}.
In the context of ultracold atoms with DDI, it has been revealed that the combined effects of TBI and DDI may lead to the formation of 
a stable supersolid state \cite {Petrov1} and a quantum droplet state \cite{Pfau1, Kui, Blakie, Chom, BoudjDp}. 
Very recently, we have shown that the TBI may shift the density profiles, the condensed fraction and the collective modes of a dipolar condensate at finite temperature \cite{Boudj4}.

Disordered dipolar Bose gases with TBI present a different physical picture and may open prospects to achieve a stable superfluid. 
The goal of this work is then to study, for the first time to the best of our knowledge, effects of a weak disorder potential with Gaussian correlation function 
on the properties of BEC with two-body interactions and TBI. 
To this end, we use the Hatree-Fock-Bogoliubov (HFB) theory which includes an additional TBI term in the momentum space. 
Our results show that the TBI are relevant in reducing the influence of the disorder potential in BEC. 
Impacts of the disorder potential and the TBI on the fluctuations, coherence and the thermodynamics of the condensate are also highlighted.
We compare our findings with previous theoretical results.

The rest of the paper is organized as follows. Section \ref{Mod} introduces the HFB model 
for a disordered dipolar Bose gases with two-body interactions and TBI. 
In Sec.\ref{GCD}, we use a correlated Gaussian disorder potential to illustrate our model and derive useful expressions for the disorder fraction and the noncondensed density.
In Sec.\ref{OBCF} we look at how the interplay of the TBI and the disorder potential enhance the coherence of the system by 
numerically  analyzing the behavior of the one-body density matrix.   
In Sec.\ref{Therm} we calculate corrections due to the disorder effects and TBI to the chemical potential and the ground state energy. 
Our conclusions are drawn in Sec.\ref{Conc}.

\section{Model} \label {Mod} 

We consider the effects of an external random potential $U(\mathbf r)$ on a dilute 3D dipolar Bose gas with contact two- and three-body interactions.
Assuming that dipoles are oriented along $z$-axis. The Hamiltonian of the system reads:
\begin{align} \label{Ham}
\nonumber \hat{H}&=\int d \mathbf r \, \hat{\psi}^\dagger (\mathbf r) \left[-\frac{\hbar^2}{2m} \Delta +U (\mathbf r) \right]\hat{\psi} (\mathbf r)\\
&+\frac{1}{2}\int d\mathbf r\int d\mathbf r' \hat{\psi}^\dagger (\mathbf r) \hat{\psi}^\dagger(\mathbf r') V(\mathbf r- \mathbf r')\hat{\psi} (\mathbf r') \hat{\psi} (\mathbf r)\\ \nonumber
&+\frac{g_{3}}{6}\int d\mathbf r\, \hat{\psi}^\dagger (\mathbf r) \hat{\psi}^\dagger (\mathbf r) \hat{\psi}^\dagger (\mathbf r) \hat{\psi} (\mathbf r) \hat{\psi} (\mathbf r) \hat{\psi} (\mathbf r),
\end{align}
where $\hat{\psi}^\dagger$ and $\hat{\psi}$ denote, respectiveiy the usual creation and annihilation field operators, $m$ is the particle mass.
The two-body interactions is described by the potential
$ V(\mathbf r- \mathbf r') =g_2\delta (\mathbf r- \mathbf r')+V_{dd} (\mathbf r -\mathbf r')$, 
where $ g_2=4\pi\hbar^2a/m $ with $a$ being the $s$-wave scattering length is assumed to be positive.
The DDI term $ V_{dd}(\mathbf r)= {\cal M}_0{\cal M}^2 (1-3 \cos^2 \theta)/4\pi r^3 $, where ${\cal M} $ is the magnetic
moment and $ \theta $ is the angle between the relative position of particles $\mathbf r$ and $z$-axis.
The three-body coupling constant $g_3$ is in general a complex number with $ Im (g_3)$ describing the three-body recombination loss and $ Re(g_3)$ 
quantifying the three-body scattering parameter.   
Here, we will assume that the imaginary part of $g_3$ is negligible  \cite {Bed, Hamid, Braa, Zhan, Blakie} 
which means that the loss rate is sufficiently small and hence, the system is stable. This well coincides with the experimental conditions reported in Ref.\cite{Pfau1}.
Note that the strength of the three-body coupling $g_3$ is related to the atomic species and can be adjusted by Feshbach resonance \cite {Kui, Evrt}.
It is therefore, hard to predict the exact value of $g_3$ (see e.g. \cite {Bulg, Braa, Braa1}).

In what follows, we suppose that the disorder potential is described by vanishing ensemble averages $\langle U(\mathbf r)\rangle=0$ and a finite disorder correlation function
$\langle U (\mathbf r) U(\mathbf r')\rangle=R (\mathbf r,\mathbf r')$. 

In the frame of the HFB formalism, the Bose-field operator can be written as 
\begin{equation} \label{FO}
\hat \psi ({\bf r},t)=\Phi ({\bf r},t)+\hat {\bar\psi} ({\bf r},t), 
\end{equation}
where $\Phi$ is the condensate wavefunction,  and $\hat {\bar\psi}$ stands for the field of the noncondensed thermal atoms.
Working in Fourier space,  the condensate
wavefunction is taken as $\Phi ({\bf r},t)= \sqrt{n_c}$ with $n_c$ being the condensate density, and 
the field operator of noncondensed atoms can be expanded in terms of 
plane waves $\hat {\bar\psi}= (1/\sqrt{V}) \sum_k a_k e^{ i \bf k. r}$. 
The DDI potential in momentum space is given by:
$V _{dd}(\mathbf k)=(\mu_0\mu^2/12 \pi) (3\cos^2\theta_{\mathbf k}-1)$, 
where the vector $\mathbf k$ represents the momentum transfer imparted by the collision.

Now we deal with a weakly interacting system and assume that the disorder is sufficiently weak.  
Then it is possible to use the Bogoliubov-Huang-Meng  approach \cite {HM} which suggests the transformation: 
\begin{align}
\hat{a}_{k} =u_k \hat{b}_k-v_k\hat{b}_{-k}^\dagger -\beta_k,\quad  \hat{a}_{k}^\dagger =u_k \hat{b}_k^\dagger-v_k\hat{b}_{-k}-\beta_k^*,
\end{align}
where $ \hat{b}_k^\dagger$ and $\hat{b}_k  $ are operators of elementary excitations,  the functions $ u_k,v_k $ are defined as
$ u_k,v_k=(\sqrt{\varepsilon_k/E_k}\pm \sqrt{E_k/\varepsilon_k})/2$ with $ E_k=\hbar^2k^2/2m $ being the free particle energy, and 
\begin{equation}
\beta_k=\sqrt{\frac{n_c}{V}}\frac{E_k}{\varepsilon_k^2} U_k.
\end{equation}
The Bogoliubov excitations energy is given by
\begin{equation}
\varepsilon_k=\sqrt{E_k^2+2n_c E_k \bar V(\mathbf k)},
\end{equation}
where $\bar V ({\bf k})= g_2 (1+g_3n_c/g_2)  [1+ \gamma (3 \cos^2 \theta_{\mathbf k}-1)]$ with $ \gamma =\epsilon_{dd}/(1+g_3n_c/g_2)$,
$\epsilon_{dd}={\cal M}_0 {\cal M}^2/(3 g_2) $ is the relative strength between the DDI and short-range interactions.
For $k \rightarrow 0$, the excitations are sound waves $ \varepsilon_k=\hbar c_s(\theta_{\bf k})k $, 
where $ c_s(\theta_{\bf k})=c_0\sqrt{(1+g_3n_c/g_2)  [1+ \gamma (3 \cos^2 \theta_{\bf k}-1)]}$ with $ c_0=\sqrt{g_2n_c/m} $ being the sound velocity without DDI and TBI.\\

The diagonal form of the Hamiltonian (\ref{Ham}) can be written as $\hat{H}=E+\displaystyle\sum_k\varepsilon_k\hat{b}_k^\dagger \hat{b}_k$.
The total energy $E=E_0 (\theta)+\delta E+E_R $, where the zeroth order term 
\begin{equation} \label{EGY0}
E_0(\theta)=\bar V (\theta) n_c N_c/2,
\end{equation}
which should be computed in the limit  $k \rightarrow 0$ since it accounts for the condensate (lowest state).
The ground-state energy shift due to quantum fluctuations is
\begin{equation} \label{EGY1}
\delta E=\frac{1}{2}\displaystyle\sum_k [\varepsilon_k-E_k-n_c \bar V ({\bf k})], 
\end{equation}
and 
\begin{equation} \label{EGY2}
E_R=-\displaystyle\sum_k  n_c \langle\mid U_k\mid^2\rangle\frac{E_k}{\varepsilon_k^2}=-\displaystyle\sum_k n_c R_k\frac{E_k}{\varepsilon_k^2},
\end{equation}
gives the correction to the ground-state energy due to the external random potential.

The noncondensed density is defined as
\begin{align} \label{NCD}
 \tilde{n}=\displaystyle\sum_k\langle\hat{a}_k^\dagger \hat{a}_k\rangle=\frac{1}{V}\displaystyle\sum_k \left[(u_k^2+v_k^2)N_k+v_k^2+\langle |\beta_{\bf k}|^2  \rangle \right],
\end{align}
%and 
%\begin{align} \label{AD}
% \tilde{m}=-\displaystyle\sum_k\langle\hat{a}_k\hat{a}_{-k}\rangle=-\frac{1}{V}\displaystyle\sum_k \left[ u_k v_k (1+2N_k)+ \langle |\beta_{\bf k}|^2  \rangle \right],
%\end{align}
where $ N_k=\langle\hat{b}_k^\dagger\hat{b}_k\rangle=1/[\exp({\varepsilon_k}/T)-1] $ is the Bose-Einstein distribution function,
and the rest of the expectation values equal to zero ($\langle\hat{b}_k^\dagger \hat{b}_k^\dagger \rangle=\langle\hat{b}_k \hat{b}_k\rangle=0$).

Inserting the expressions of $u_k$ and $v_k$ in Eq.(\ref{NCD}), and  working in the thermodynamic limit 
where the sum over $k$ can be replaced by the integral $\sum_{\bf k}=V\int d \mathbf k/(2\pi)^3$, we get 
\begin{subequations} \label{NCDD}
\begin{align} 
\tilde n&= \tilde n_0+\tilde n_{th} +n_R, \\
&=\frac{1}{2}\int \frac{d \mathbf k} {(2\pi)^3} \left[\frac{E_k+\bar V(\mathbf k) n_c} {\varepsilon_k} -1\right] \label{NCD1} \\
&+\frac{1}{2}\int \frac{d \mathbf k} {(2\pi)^3} \frac{E_k+\bar V(\mathbf k) n_c} {\varepsilon_k}\left[\coth\left(\frac{\varepsilon_k}{2T}\right)-1\right] \label{NCD2}  \\
&+n_c\int \frac{d \mathbf k} {(2\pi)^3} R_k \frac{ E_k^2}{\varepsilon_k^4}. \label{NCD3}
\end{align}
\end{subequations}
The leading term (\ref{NCD1}) denotes the zero temperature contribution to the noncondensed density.
The subleading term (\ref{NCD2}) stands for thermal fluctuation corrections to the noncondensed density.
Whereas the third term (\ref{NCD3}) represents the condensate fluctuations due to the disorder potential known as {\it glassy fraction} and originates from 
the accumulation of density near the potential minima and density depletion around the maxima.

%The anomalous density reads
%\begin{subequations} \label{ADD}
%\begin{align} \label{AD0}
%\tilde m &= \tilde m_0+\tilde m_{th} +n_R, \\
%&-\frac{1}{2}\int \frac{d \mathbf k} {(2\pi)^3} \frac{ \bar V (\mathbf k) n_c} {\varepsilon_k}  \label{AD1}\\
%&-\frac{1}{2}\int \frac{d \mathbf k} {(2\pi)^3} \frac{ \bar V (\mathbf k) n_c} {\varepsilon_k}\coth\left(\frac{\varepsilon_k}{2T}\right)  \label{AD2} \\
%&+n_c\int \frac{d \mathbf k} {(2\pi)^3} R_k \frac{ E_k^2}{\varepsilon_k^4}. 
%\end{align}
%\end{subequations}
%The zero temperature term (\ref{AD1}) in the anomalous density is ultraviolet divergent.
%This divergency comes from the contact interactions. To overcome such a problem one should 
%use the dimensional regularization which is valid for very dilute gases, and gives for the integral $\int_0^{\infty} dx (x/ \sqrt{1+x^2})=-1$ \cite{Anders, Yuk, Boudjbook}.
%The second term (\ref{AD2}) accounts for thermal contributions to the anomalous density.
%The glassy fraction $n_R$ already defined in (\ref{NCD3}).

\section{Gaussian-correlated disorder } \label{GCD}

As a concrete example, we consider in this section the case of a correlated Gaussian disorder model, which 
allow for unique control of the interplay between the disorder potential and interactions in both dipolar and nondipolar BEC \cite{Krum, Boudj1}. 
It can be written as
\begin{equation} \label {Gdis}
R(k)=R_0\exp[-\sigma^2k^2/2],
\end{equation} 
where $ R_0 $ is the disorder strength which has dimension (energy)$^2 \times$(length)$^3 $ and $ \sigma$
characterizes the correlation length of the disorder.

The glassy fraction can be calculated easily via Eq.(\ref{NCD3})
\begin{equation}
n_R=n_{\text{HM}}  (1+g_3n_c/g_2)^{-1/2} h(\gamma,\sigma/\xi),
\end{equation}
where $n_{\text{HM}}=[m^2 R_0/8\pi^{3/2}\hbar^4]\sqrt{n_c/a} $ is the usual Huang-Meng result \cite{HM}. 
The anisotropic disorder function is given as
\begin{align} \label{HF}
h (\gamma,\sigma/\xi)=\int^\pi_0 \mathrm{d}\theta\frac{\sin\theta S(\alpha)}{2\sqrt{1+\gamma(3\cos^2\theta-1)}},
\end{align}
where the function $ S(\alpha)=e^{2\alpha} (4\alpha+1) \left[1-\text{erf}(\sqrt{2\alpha})\right]-2\sqrt{2\alpha/\pi}$, 
and $\alpha=\sigma^2[\epsilon_{dd}/\gamma (1+\gamma(3\cos^2\theta-1))]/\xi^2 $ with $ \xi=\hbar/\sqrt{mn_cg_2} $ being the healing length.
In the absence of the DDI ($\epsilon_{dd}=0$), and in the limit $ \sigma/\xi \rightarrow 0 $ and $g_3=0 $, one has $ h(\gamma,\alpha)\rightarrow 1$, thus,
one recovers the well-known Hang and Meng result ($n_R=n_{\text{HM}}$) \cite{HM}.

\begin{figure}
\includegraphics[scale=0.45]{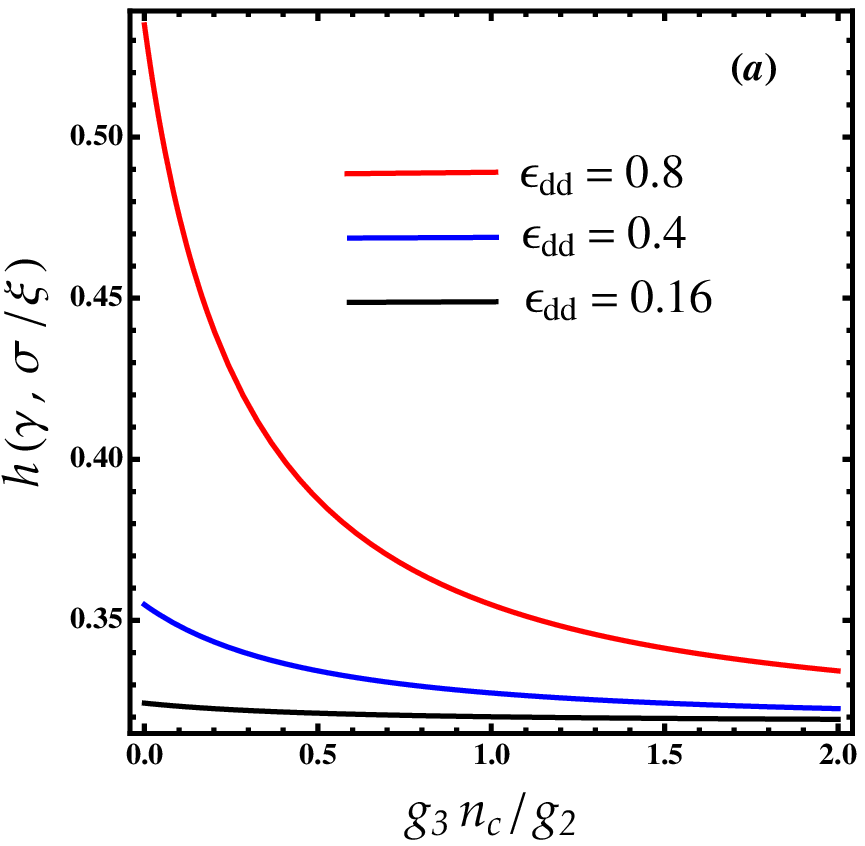}
\includegraphics[scale=0.45]{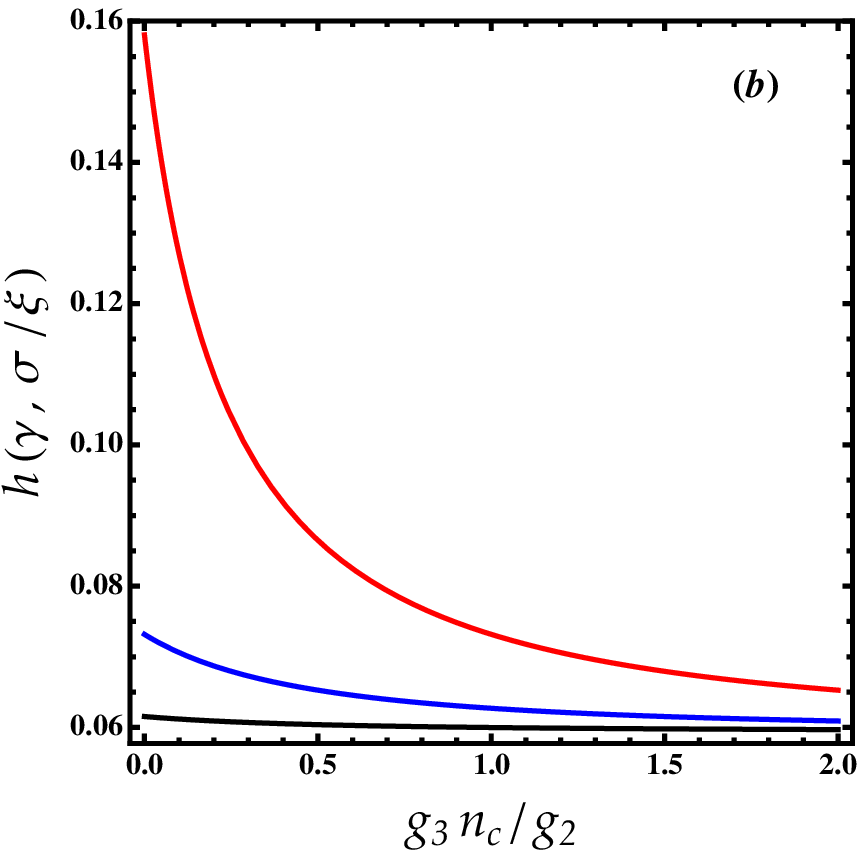}
\caption{ Disorder function  $ h(\gamma, \sigma/\xi) $, as a function of $g_3 n_c/g_2$ for several values of $\epsilon_{dd}$ for $\sigma/\xi=0.4$ (a) 
and $\sigma/\xi=1.2$ (b).}
\label{DisF}
\end{figure}

The effects of both correlation length and effective interaction parameter $\gamma$ on the behavior of the disorder function are presented in Fig.\ref{DisF}.
We observe that the function $h(\gamma,\alpha)$ is decreasing with $g_3 n_c/g_2$ indicating that
the TBI lead to reduce the disorder fluctuations (glassy fraction) inside the condensate even in the limit $ \sigma<\xi $. 
As is expected, the disorder fraction becomes significant for large DDI in contrast to the case of a disordered dipolar BEC with Lee-Huang-Yang (LHY) quantum corrections \cite{Boudj5}.
The main difference between the TBI and the LHY corrections is that these latter are valid only in the regime of  weak disorder since they are computed
within the local density approximation which assumes that the external random potential should vary smoothly in space on a length scale comparable to the healing length 
or the characteristic correlation length of the disorder \cite{Boudj5}.  Whereas, the TBI still remain applicable for both weak and strong disorder potentials.
For $ \sigma >\xi $, the disorder effects is not important (see Fig.\ref{DisF}.b). 

For delta-correlated disorder where $ \sigma/\xi \rightarrow 0 $,  the function $ h(\gamma,0) ={\cal Q}_{-1}(\gamma)$ and 
$ n_R=n_{HM} {\cal Q}_{-1}(\gamma) $, where the contribution of the DDI is expressed by the functions 
${\cal Q}_j (x)=\int_0^1 dy (1-x+3xy^2)^{j/2}$ \cite{Krum, Ghab, Boudj, Boudj1,Boudj5}. 
Note that the functions ${\cal Q}_j (x)$ tend to unity for $ \gamma=0 $ ($ {\cal Q}_j(0)=1$), and become imaginary for $ \gamma>0 $.

Now, we focus ourselves to calculate quantum and thermal depletion in a disordered BEC. 
Integrals (\ref {NCD1}) and (\ref {NCD2})  yield, respectively
\begin{align} \label{NDQ1}
\frac{\tilde{n}_0}{n_c}=\frac{8}{3}\sqrt{\frac{n_ca^3}{\pi}}  (1+g_3n_c/g_2)^{3/2}  {\cal Q}_{3}(\gamma),
\end{align}
and 
\begin{align} \label{NDT1}
\frac{\tilde{n}_{th}}{n_c}=\frac{2}{3} \left(\frac{\pi T}{n_c g_2} \right)^2\sqrt{\frac{n_ca^3}{\pi}}   (1+g_3n_c/g_2)^{-1/2} {\cal Q}_{-1}(\gamma).
\end{align}
%and 
%  \begin{align} \label {ADD1}
%  \nonumber \frac{\tilde{m}}{n_c}=8\sqrt{\frac{n_ca^3}{\pi}} \left(\frac{\epsilon_{dd}}{\gamma} \right)^{3/2}  {\cal Q}_{3}(\gamma)
%+\frac{n_{HM}}{n_c} h(\gamma,\alpha)\\-\frac{2}{3} \left(\frac{\pi T}{n_cg_2} \right)^2\sqrt{\frac{n_ca^3}{\pi}} \sqrt{\frac{\gamma}{\epsilon_{dd}}}  {\cal Q}_{-1}(\gamma).
%  \end{align}
For $\epsilon_{dd}=0$ and $g_3=0$, we recover the standard expressions for $\tilde{n}_0$ and $\tilde{n}_{th}$ .
When $g_3=0$, Eqs.(\ref{NDQ1}) and (\ref{NDT1}) reduce to that obtained in our previous work for a dipolar BEC without TBI \cite{Boudj1}.

\section{One-body density matrix } \label{OBCF}

\begin{figure}
\begin{center}
 \includegraphics[scale=0.45]{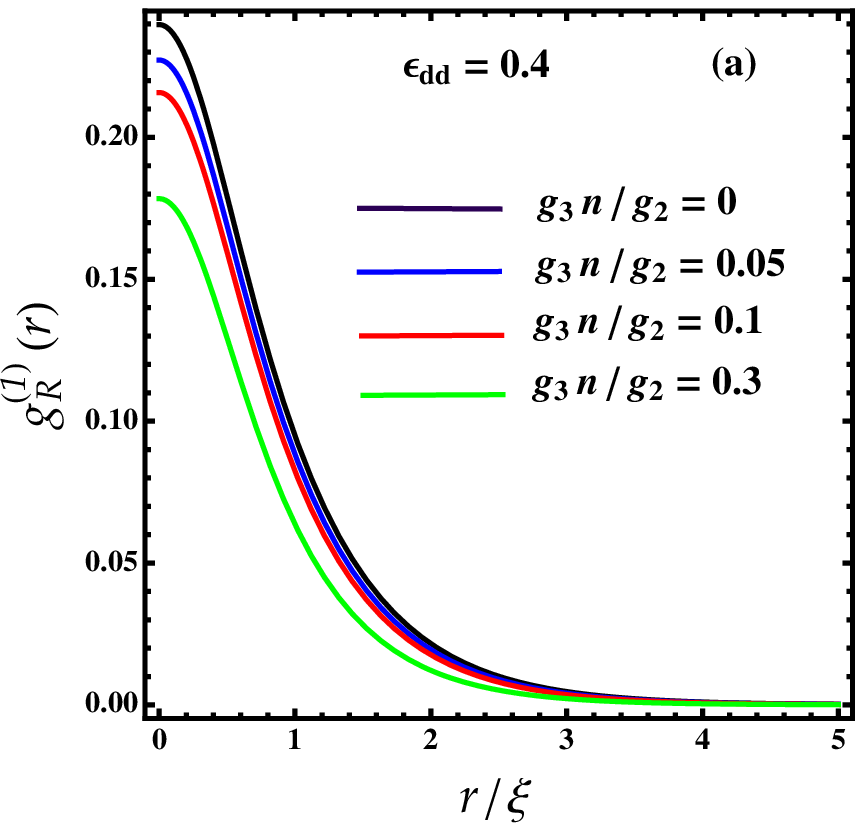}
 \includegraphics[scale=0.45]{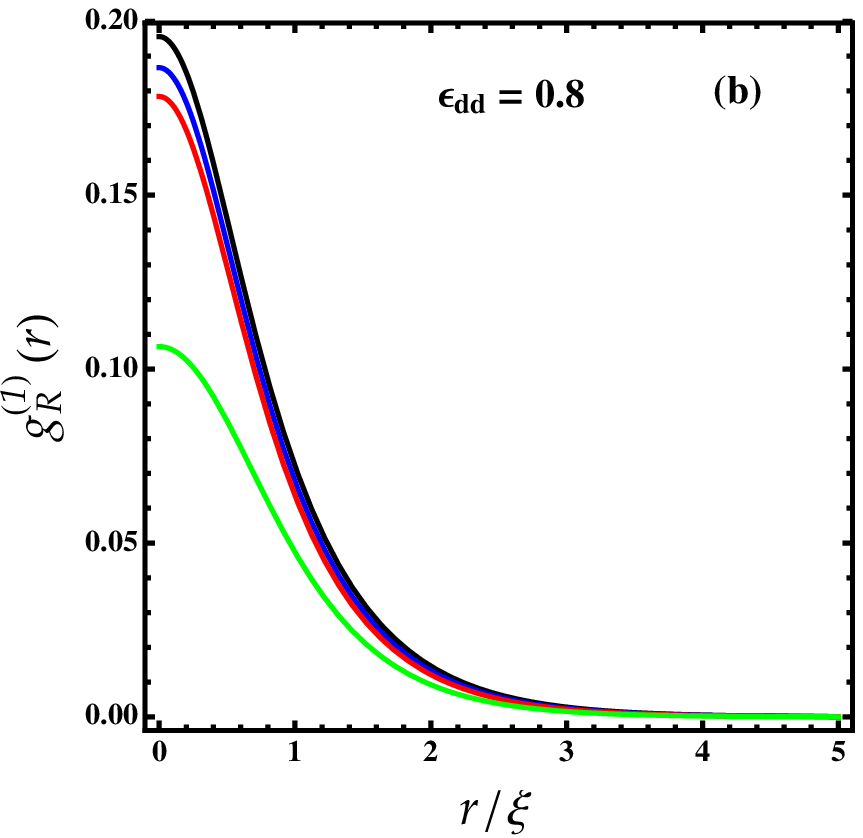}
 \includegraphics[scale=0.45]{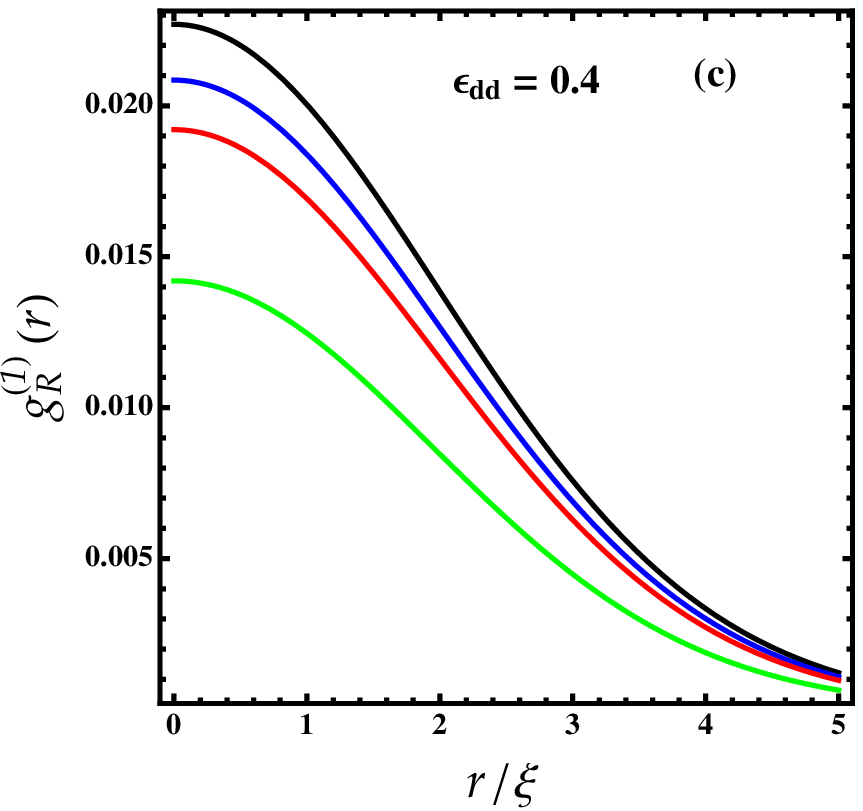}
 \includegraphics[scale=0.45]{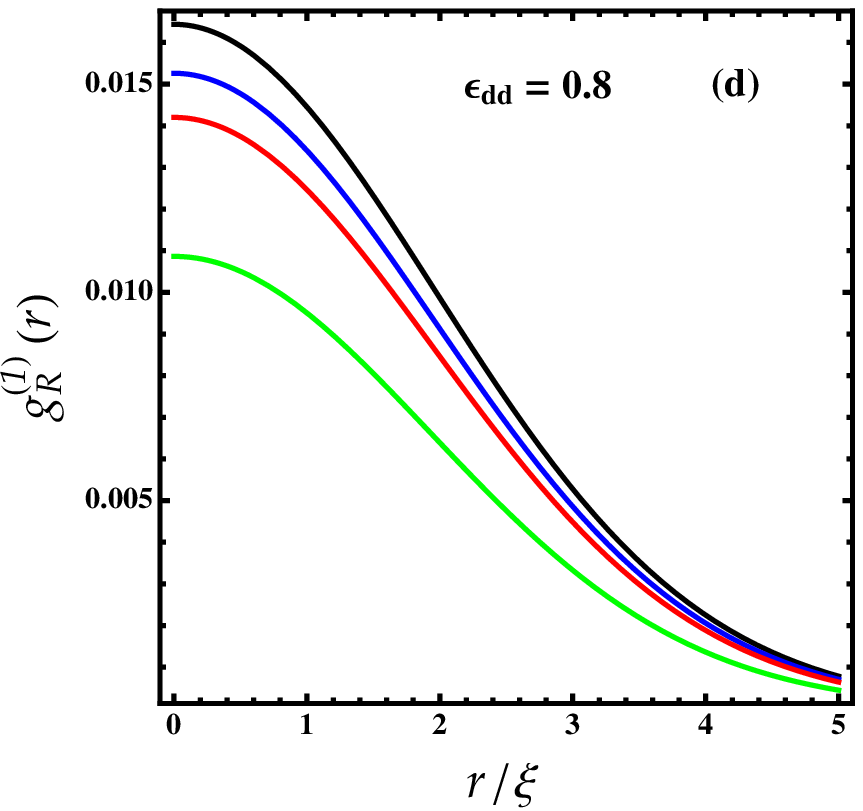}
\end{center} 
\caption{(color online) One-body density matrix  due to the disorder corrections, $\mathrm{g}_R^{(1)}(r) $, for
$\sigma/\xi=0.2$ (a)-(b) and $\sigma/\xi=1.2$ (c)-(d).}
\label{corr}
\end{figure} 

The one-body density matrix (first-order correlation function) is defined as $g^{(1)}(\mathbf {r},\mathbf{r'},t,t')=\langle \hat\psi^\dagger (\mathbf r,t) \hat \psi(\mathbf r',t')\rangle$.
In uniform case it depends only on the difference $ \vert{\bf r-r'} \vert=r$.
Using the decomposition (\ref{FO}), expressing the noncondensed field operator as 
$\hat {\bar\psi}=(1/V) \sum_k [u_k \hat b_k e^{i \bf k. r}- v_k\hat b_k^\dagger e^{-i \bf k. r}]$,
and then taking into account that $|\Phi ({\bf r},t)|=\sqrt{n_c}$. We thus, get
\begin{align} \label{corrF}
g^{(1)} ({\bf r})&=n_c+g_R^{(1)}({\bf r}) +\int_0^{\infty}  \frac{d \mathbf k}{(2\pi)^d}\left[v_{k}^2 + (u_{k}^2+v_{k}^2)N_{k}\right] e^{i {\bf k \cdot r}},
\end{align} 
The second term $g_R^{(1)}({\bf r})= \int (d \mathbf k/(2\pi)^3)  \langle |\beta_{\bf k}|^2  \rangle\, \mathrm{e}^{i\bf {k.r}}$
represents the disorder effects on the first order correlation function. The  behavior of  $g_R^{(1)}({\bf r})$ is displayed in Fig.\ref{corr}.
We observe that for small disorder correlation length ($\sigma/\xi =0.2$),  $g_R^{(1)}(r) $ is decreasing with increasing the TBI and the DDI (see Fig.\ref{corr}. a-b ).
The same behavior holds for large $\sigma$. 
Importantly, $g_R^{(1)}(r) $ vanishes at large distance $r$ in both cases signaling the non-existence of
mini condensates formed by the localized particles in  the respective minima of the external random potential. 
This does not mean	that the long-range order of the whole system is destroyed.			

The last term in Eq.(\ref{corrF}) accounts for the quantum and thermal contributions to the one-body correlation function.
One can easily show that this term decays at $r \rightarrow \infty$ and thus, $g^{(1)}({\bf r})$ tends to $n_c$,
revealing the existence of the long-range order (true condensate).
Note that the DDI, the TBI and the temperature can also shift  the one-body correlation function.

\section {Thermodynamic quantities} \label{Therm}

In this section, we calculate disorder corrections to some thermodynamic quantities such as the chemical potential and the ground state energy.\\
Within the realm of the HFB theory, the chemical potential can be written as 
 \begin{equation} \label{EoS}
 \mu= \mu_0+\delta\mu+2\mu_R,
 \end{equation}
 where 
 \begin{equation} \label{EoS1}
\mu_0=\bar V(0) n_c,
 \end{equation}
is the first-order chemical potential \cite{Boudj6}.\\
Corrections to the chemical potential due to the disorder effects are given as 
 \begin{equation} \label{EoS2}
 \mu_R= g_2 n_{\text{HM}}  (1+g_3n_c/g_2)^{1/2} H (\gamma,\sigma/\xi),
 \end{equation}
where 
\begin{align}  
 H (\gamma,\sigma/\xi)=\frac{1}{2} \int^\pi_0 \mathrm{d}\theta\sin\theta\sqrt{1+\gamma(3\cos^2\theta-1)}S(\alpha),
\end{align} 
Corrections to the chemical potential due to the quantum and thermal fluctuations are defined as :
$\delta \mu=\sum\limits_{\bf k} \bar V (\mathbf k) \left[v_k(v_k-u_k)+(v_k-u_k)^2N_k \right]$ \cite{BoudjDp, Boudj6}.
Nevertheless, this chemical potential cannot be evaluated straightforwardly since the zero-temperature term is ultraviolet divergent.
Such a problem can be worked out either by using the dimensional regularization which is valid for very dilute gases \cite{Anders, Yuk, Boudjbook}
or by renormalizing the contact interaction through the $T-$matrix approach \cite {FV}.
After some algebra, the resulting corrections to the chemical potential read
 \begin{align} \label{EoS3}
\frac{\delta\mu}{g_2 n_c}&=\frac{32}{3} \sqrt{\frac{n_ca^3}{\pi}}   (1+g_3n_c/g_2)^{5/2} {\cal Q}_5 (\gamma) \nonumber\\
&+\frac{2}{3} \left(\frac{\pi T}{n_cg_2} \right)^2\sqrt{\frac{n_c a^3}{\pi}}   (1+g_3n_c/g_2)^{1/2} {\cal Q}_{1}(\gamma) .
 \end{align}
\begin{figure}
 \includegraphics[scale=0.8]{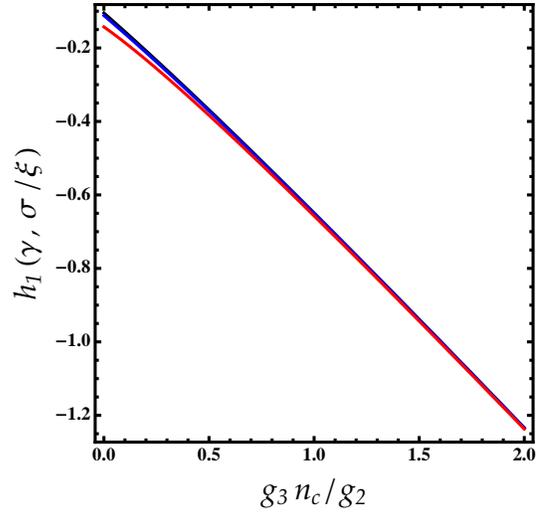}
\caption{(Color online) Disorder energy function  $ h_1(\gamma,\sigma/\xi) $ versus $ g_3n_c/g_2 $ for $\sigma/\xi =0.5$. 
Black line: $ \epsilon_{dd}=0.16 $, blue line: $ \epsilon_{dd}=0.4 $, and red line: $  \epsilon_{dd}=0.8 $.}
\label{Efunc}
\end{figure} 
Importantly, for $g_3=0$, the total chemical potential (\ref{EoS}) reduces to that obtained in our recent  work \cite{Boudj6}. 
For a cleaned ($R_0=0$) condensate with two-body contact interactions ($g_3= \epsilon_{dd}=0$), 
the obtained corrections to the chemical potential coincide  with the seminal Lee-Huang-Yang quantum corrected equation of state \cite{LHY}.\\

The energy shift  (\ref{EGY2}) due to the disorder effects is finite and it can be evaluated as
\begin{equation}
 \frac{E_R}{N}= \frac{2mR_0}{\hbar^2} (1+g_3n_c/g_2)^{1/2} \sqrt{\frac{n_c a }{\pi}} {\cal H} (\gamma,\sigma/\xi),
\end{equation}
where the function
\begin{align} 
{\cal H} (\gamma,\sigma/\xi)=\frac{1}{2} \int^\pi_0 \mathrm{d}\theta\sin\theta\sqrt{1+\gamma(3\cos^2\theta-1)}S_1(\alpha),
\end{align} 
and the function $ S_1(\alpha)=e^{2\alpha} \text{erfc} (\sqrt{2\alpha})-\sqrt{1/2\alpha} $.
The disorder energy function  $ h_1(\gamma,\sigma/\xi) $ is decreasing with $ g_3 $ as is seen in Fig.\ref{Efunc} indicating that
the TBI lead to lower the energy due to the disorder fluctuations which is in agreement with the above results.
We observe also that for $ g_3n_c/g_2 \leq 0.7 $, the DDI effects on the energy are more pronounced.

Corrections to the energy due to the quantum and thermal fluctuations can be calculated easily through Eq.(\ref{EGY1}) or by integrating the chemical potential 
(\ref{EoS3}) with respect to the density.

\section{Conclusion} \label{Conc}

In this paper, we investigated the properties of dipolar Bose gas with TBI subjected to a correlated Gaussian disorder. 
We showed that the DDI may lead to arrest transport of atoms under disorder augmenting the glassy fraction inside the condensate, while
the presence of the TBI may lead to a diffusive motion of particles.
We pointed out that the one-body density matrix is a decreasing function with the TBI. 
We calculated in addition the chemical potential of a disordered dipolar BEC and ultraviolet divergences are removed by means of dimensional regularization.
The combined effects of the DDI, TBI, and temperature found to crucially affect the chemical potential and the ground state energy of the system.
%The role of the TBI in the Anderson localization will be an interesting topic for a future work.

 \section*{Author Contributions}  
All authors discussed the results and made critical contributions to the work.
AB contributed to the writing of the manuscript.

\end{document}